\documentstyle[prd,aps,floats,epsfig]{revtex}
\renewcommand{\gg}{\gamma} 
\newcommand{\gd}{\delta} 
\renewcommand{\ge}{\epsilon} 
\newcommand{\gve}{\varepsilon}

\newcommand{\gk}{\kappa} 
\newcommand{\gl}{\lambda} 
\newcommand{\gr}{\rho} 
\newcommand{\gth}{\theta} 
\newcommand{\gs}{\sigma} 
 
\newcommand{\go}{\omega}

\newcommand{\gps}{\psi}

\newcommand{\gF}{\Phi}

\newcommand{\gL}{\Lambda} 
 
\newcommand{\gTh}{\Theta} 
\newcommand{\gO}{\Omega}

\newcommand{\cH}{{\cal H}}

\newcommand{\cL}{{\cal L}}

\newcommand{\bQ}{{\bar Q}}

\newcommand{\bge}{{\bar\epsilon}}

\newcommand{\bgps}{{\bar\psi}}

\newcommand{\bgF}{{\bar\Phi}}

\newcommand{\slashed}{\hspace{-1.1ex}/} 
\newcommand{\Slashed}{\hspace{-1.7ex}/\hspace{.6ex}}

\newcommand{\der}{\partial}

\newcommand{\sder}{\der\slashed}

\newcommand{\nit}{\noindent} 
\newcommand{\nl}{\newline} 
 
\newcommand{\dsp}{\displaystyle}

\newcommand{\ct}{\cite} 
\newcommand{\bit}{\bibitem} 
\newcommand{\lh}{\left(} 
\newcommand{\rh}{\right)}

\newcommand{\hs}[1]{\hspace{#1 em}}
 
\newcommand{\beq}{\begin{equation}} 
\newcommand{\feq}{\end{equation}} 
\newcommand{\barr}{\begin{array}} 
\newcommand{\earr}{\end{array}}

\begin{document}
\draft

\title{Superhydrodynamics}
\author{T.S.\ Nyawelo$^a$\thanks{\tt tinosn@nikhef.nl}, 
J.W.\ van Holten$^a$\thanks{\tt v.holten@nikhef.nl},
S.\ Groot Nibbelink$^{b}$\thanks{\tt nibblink@th.physik.uni-bonn.de} 
}
\address{$^a$ NIKHEF, P.O.\ Box 41882, Amsterdam NL 
\\$^b$ Phys.\ Institut, Bonn University, Nussallee 12 D-53115 Bonn BRD}

\maketitle

\begin{abstract} 
We present a covariant and supersymmetric theory of relativistic 
hydrodynamics in four-dimensional Minkowski space. 
\end{abstract}

\vskip1cm

\nit
Relativistic fluid mechanics \ct{LL} is generally simpler to formulate than  
its non-relativistic counterpart. As it is also believed to provide a more 
accurate description of hydrodynamical phenomena, much work has been 
invested in its development \ct{bcart,lebkhal}. Recently, an interesting 
extension of the theory to include non-abelian charges and currents has 
been proposed \ct{jack1}. One of the important aspects of this formalism 
is that it includes vorticity consistently at the lagrangean level, by 
developing a non-abelian generalization of the Clebsch decomposition of 
the vector conjugate to the current; for a review with many references, see 
\ct{jack}. 

In a related development, Jackiw and Polychronakos \ct{jack2} have presented 
a supersymmetric theory of fluid mechanics in (2+1)-dimensional space-time.
This model is rather special, as it descents from a supermembrane theory 
in (3+1) dimensions \ct{jack3,hoppe,horv}. It results in a supersymmetric 
generalization of the non-relativistic planar Chaplygin gas \ct{chap}.
An interesting result obtained in \ct{jack2} is, that the vorticity in the 
theory is generated by the fermion fields, rather then by the bosonic component
of the fluid. In spite of these advances, so far a relativistic and 
supersymmetric theory of fluid mechanics in (3+1) dimensions is lacking. 
The present work contributes to filling this gap. 

The main result of this work is a supersymmetric component action, which we
present both in its lagrangean and hamiltonian form, for a pseudo-classical 
conserved current $V_{\mu}$ made up of bosonic as well as fermionic 
contributions. To be more precise, the current is constructed from two real 
scalar fields $(C,N)$ and a chiral fermion field $\gps_{\pm}$\footnote{Our 
conventions for chiral spinors are such, that $\gg_5 \gps_{\pm} = \pm 
\gps_{\pm}$ and $\bgps_{\pm} \gg_5 = \pm \bgps_{\pm}$; charge conjugations 
acts as $\gps_{\pm} = C \bgps_{\pm}^T$, where $\bgps_{\pm} = i 
\gps_{\mp}^{\dagger} \gg_0$. It should also be noted, that the euclidean 
$\gg_ 4 = i \gg_0$ is hermitean, hence $\gg_0$ is anti-hermitean.}: 
\beq 
V_{\mu} = \frac{1}{G(C)}\, \lh \der_{\mu} N + \frac{i}{2}\, G^{\prime}(C)\, 
 \bgps_+ \gg_{\mu} \gps_- \rh, \hs{2} \der \cdot V = 0.  
\label{1}
\feq 
Here $G(C)$ is some function of the real scalar field $C$ which, for the 
purpose of constructing models, needs no further specification. However,
under particular conditions discussed below the current allows an elegant 
and straightforward interpretation in terms of fluid flow. 

The fields $(C,N,\gps_{\pm})$ and $V_{\mu}$ representing the conserved current 
are described by the supersymmetric lagrangean: 
\beq 
\cL = - \frac{1}{2}\, G(C) \left[ \lh \der_{\mu} C \rh^2 
 - V_{\mu}^2 + \bgps_+ \stackrel{\leftrightarrow}{\sder} \gps_- \right] -\, 
 V^{\mu} \lh \der_{\mu} N + \frac{i}{2}\, G^{\prime}(C) \bgps_+ 
 \gg_{\mu} \gps_- \rh - \frac{1}{8}\, G^{\,\prime\prime}(C)\, 
 \bgps_+ \gps_+ \bgps_- 
 \gps_-.    
\label{2}
\feq 
The infinitesimal supersymmetry transformations leaving the action invariant, 
parametrized by anti-commuting spinor parameters $\ge_{\pm}$, are:
\beq 
\barr{ll} 
\gd C = \bge_+ \gps_+ + \bge_- \gps_-, & \dsp{ \gd N = i G(C)\, 
 \lh \bge_- \gps_- - \bge_+ \gps_+ \rh, }\\
 & \\
\gd \gps_+ = \lh \sder C + i V\Slashed \rh \ge_-, & 
\gd \gps_- = \lh \sder C - i V\Slashed \rh \ge_+, \\ 
 & \\  
\gd V_{\mu} = 2 i \bge_+ \gs_{\mu\nu} \der^{\nu} \gps_+ - 2 i \bge_- 
 \gs_{\mu\nu} \der^{\nu} \gps_-, &
\earr
\label{3}
\feq 
Under these transformations the variation of the lagrangean is a total 
divergence: 
\beq 
\gd \cL = \der^{\mu} \lh \bge_+ B_{+\mu} + \bge_- B_{-\mu} \rh, 
\label{4}
\feq 
where the vector-spinor fields $B_{\pm \mu}$ are given, modulo equations 
of motion
, by  
\beq 
\barr{l} 
B_{+\mu} \simeq \dsp{ -\frac{1}{2}\, G(C)\, \gg_{\mu}\,
 (\sder C - i V\Slashed)\, \gps_+ - \frac{1}{2}\, 
 G^{\,\prime}(C)\, \gg_{\mu} \gps_- \bgps_+ \gps_+, }\\
 \\
B_{-\mu} \simeq \dsp{ -\frac{1}{2}\, G(C)\, \gg_{\mu}\,  
 (\sder C + i V\Slashed)\, \gps_- - \frac{1}{2}\, 
 G^{\,\prime}(C)\, \gg_{\mu} \gps_+ \bgps_- \gps_-. }
\earr 
\label{5} 
\feq 
The commutator of two supersymmetry transformations (\ref{3}) closes as 
usual, modulo field equations. A complete and manifest off-shell formulation 
can be obtained from a superspace treatment. The relevant superspace action 
is defined in terms of a real vector superfield $V$ with lowest components 
$(C, \gps_{\pm},V_{\mu})$, and a pair of conjugate chiral superfields 
$(\gL_+, \gL_-)$ with lowest component $z$ such that Re$\,z = N$. The 
superfield expression for the lagrangean is 
\beq 
S\, =\, \int d^4 x\, \int d^2 \gth_+ \int d^2 \gth_-\, 
 \lh \frac{1}{2}\, V \lh \gL_+ + \gL_- \rh - F(V) \rh. 
\label{6}
\feq 
Here $F(V)$ is an arbitrary real function of the superfield $V$, with lowest 
scalar component $F(V)|_{\gth_{\pm} = 0} = F(C)$. Reduction of the superspace 
expression (\ref{6}) in terms of components leads, after elimination of a 
number of auxiliary fields and some rescaling of the scalars, to the 
expression (\ref{2}) with the identification $G(C) = F^{\prime\prime}(C)$. 

\nit
The field equations derived from the lagrangean (\ref{2}) are: 
\beq 
\barr{l} 
\der \cdot V = 0, \hs{2} \dsp{ \der_{\mu} N = G(C) V_{\mu} 
 - \frac{i}{2}\, G^{\prime}(C)\, \bgps_+ \gg_{\mu} \gps_-, }\\
 \\ 
G(C) \Box\, C = \dsp{ - \frac{1}{2}\, 
 G^{\,\prime}(C) \left[ (\der C)^2 + V^2 - 
 \bgps_+ \stackrel{\leftrightarrow}{\sder} \gps_- \right] + \frac{i}{2}\, 
 G^{\,\prime\prime}(C)\, \bgps_+ V\Slashed \gps_- + \frac{1}{8}\, 
 G^{\,\prime\prime\prime}(C)\, \bgps_+ \gps_+ \bgps_- \gps_-, } 
\earr
\label{7}
\feq
for the bosonic fields, and 
\beq
\barr{l}
G(C)\, \sder \gps_{\pm} = \dsp{ - \frac{1}{2} 
 G^{\prime}(C) \lh \sder C \mp i V\Slashed \rh \gps_{\pm} 
 - \frac{1}{4}\, G^{\prime\prime}(C)\, \gps_{\mp} \bgps_{\pm} \gps_{\pm}, }
\earr 
\label{8}
\feq 
for the fermionic ones.  The first two equations (\ref{7}) indeed reproduce
equation (\ref{1}). 

Because of translation invariance and supersymmetry, the theory described 
by $\cL$ conserves four-momentum and supercharge. The corresponding 
currents are provided by the energy-momentum tensor and the supercurrents. 
For the energy-momentum tensor we use the symmetrized version 
\beq 
\barr{l} 
T_{\mu\nu} = \dsp{ G(C) \left[ \der_{\mu} C\, \der_{\nu} C 
+ V_{\mu} V_{\nu} + \frac{1}{4}\, \bgps_+ \lh \gg_{\mu} 
 \stackrel{\leftrightarrow}{\der}_{\nu} + \gg_{\nu} 
 \stackrel{\leftrightarrow}{\der}_{\mu} \rh \gps_- \right] - \frac{i}{4}\, 
 G^{\,\prime}(C)\, \bgps_+ \lh \gg_{\mu} V_{\nu} + \gg_{\nu} 
 V_{\mu} \rh \gps_- }\\
 \\ 
\dsp{ \hs{2.6} - g_{\mu\nu} \left[ \frac{1}{2}\, G(C) 
 \lh (\der C)^2 + V^2 + \bgps_+ \stackrel{\leftrightarrow}{\sder} 
 \gps_- \rh + \frac{1}{8}\, G^{\,\prime\prime}(C)\, 
 \bgps_+ \gps_+ \bgps_- \gps_- \right]. }
\earr
\label{9}
\feq 
It is obtained from the non-symmetric translational Noether currents 
$\gTh_{\mu\nu}$ by the addition of an improvement term: $T_{\mu\nu} =
\gTh_{\mu\nu} + \gO_{\mu\nu}$, where 
\beq 
\gO_{\mu\nu} = \frac{1}{4}\, \der^{\gl} \lh \gve_{\mu\nu\gk\gl} 
 G(C)\, \bgps_+ \gg_{\gk} \gps_- \rh, \hs{3} \der^{\mu} \gO_{\mu\nu} = 0.
\label{10}
\feq 
The field equations (\ref{7}) and (\ref{8}) imply $\der^{\mu} T_{\mu\nu} = 0$,
from which the conservation of four-momentum $P_{\mu} = \int d^3 x T_{\mu 0}$ 
follows. The supercurrents are obtained directly by Noether's procedure and 
read (in an on-shell version)
\beq 
\barr{l} 
S_{+\mu} = \dsp{ G(C)\, (\sder C - i V\Slashed)\, \gg_{\mu} 
 \gps_+ - \frac{1}{2}\, G^{\,\prime}(C)\, \gg_{\mu} \gps_-\, \bgps_+ \gps_+, }\\
 \\
S_{-\mu} = \dsp{ G(C)\, (\sder C + i V\Slashed)\, \gg_{\mu} 
 \gps_- - \frac{1}{2}\, G^{\,\prime}(C)\, \gg_{\mu} \gps_+\, \bgps_- \gps_-. }
\earr
\label{11}
\feq 
As on shell $\der \cdot S_{\pm} = 0$, the conservation of the supercharges 
$Q_{\pm} = \int d^3x S_{\pm 0}$ follows. 

The manifestly covariant lagrangean description of the theory has an equivalent
canonical formulation in terms of a hamiltonian with a corresponding bracket
structure. First, the canonical momenta are defined by 
\beq 
\barr{ll} 
\dsp{ \pi_C = \frac{\gd \cL_{eff}}{\gd \dot{C}}\, = G(C)\, 
 \dot{C}, }& \dsp{ \pi_N = \frac{\gd \cL_{eff}}{\gd \dot{N}}\, = V_0, }\\
 & \\
\dsp{ \pi_+ = \gg_0\, \frac{\gd \cL_{eff}}{\gd \dot{\bgps}_-}\, = 
 \frac{1}{2}\, G(C)\, \gps_+, }& 
\dsp{ \pi_- = \gg_0\, \frac{\gd \cL_{eff}}{\gd \dot{\bgps}_+}\, =
 \frac{1}{2}\, G(C)\, \gps_-. }
\earr
\label{12}
\feq 
The inclusion of the $\gg_0$ in the definition of the fermionic momenta is 
motivated by the general requirement that the charge conjugation properties 
of the momenta reflect those of the fermion variables themselves. Furthermore 
note, that $V_0$ is a canonical momentum, whereas the 3-vector field $\vec{V}$ 
is an auxiliary field, which can be eliminated by its algebraic field equation; 
in particular in the following we use the identifications
\beq 
V_0 = \pi_N, \hs{2} 
\vec{V} = \frac{1}{G(C)}\, \lh \vec{\nabla} N + 
 \frac{i}{2}\, G^{\prime}(C)\, \bgps_+ \vec{\gg}\, \gps_- \rh. 
\label{13}
\feq 
As usual, the fermionic momenta are not independend, and the theory possesses 
the second-class constraints: 
\beq 
\chi_{\pm} = \pi_{\pm} - \frac{1}{2}\, G(C) \gps_{\pm} 
 \simeq 0. 
\label{14}
\feq 
As a result, the dynamics of the theory is generated by a hamiltonian 
---obtained from $\cL$ by a straightforward Legendre transformation--- and a 
set of Dirac-Poisson brackets derived by reduction of the standard Poisson 
brackets to the physical shell (\ref{14}) in the full phase-space. In summary, 
one first constructs a hamiltonian density
\beq 
\barr{l} 
\cH = \dsp{ \frac{1}{2G(C)}\, \pi_C^2 + \frac{1}{2}\, G(C)\, \pi_N^2 - 
 \frac{i}{2}\, G^{\prime}(C)\, \pi_N\, \bgps_+ \gg_0 \gps_- + \frac{1}{2}\, 
 G(C) \lh (\vec{\nabla} C)^2 + \bgps_+ 
 \stackrel{\leftrightarrow}{\nabla\Slashed} \gps_- \rh }\\ 
 \\
\dsp{ \hs{2.2} +\, \frac{1}{2G(C)}\, \lh \vec{\nabla} N + \frac{i}{2}\, 
 G^{\prime}(C)\, \bgps_+ \vec{\gg}\, \gps_- \rh^2 + \frac{1}{8}\, 
 G^{\prime\prime}(C)\, \bgps_+ \gps_+ \bgps_- \gps_-, }
\earr
\label{15}
\feq 
where it is to be noted, that $\nabla\Slashed = \vec{\gg} \cdot \vec{\nabla}$ 
now denotes a 3-dimensional contraction. Subsequently the field equations are 
obtained from the hamiltonian $H = \int d^3x \cH$ by the Dirac-Poisson 
brackets 
\beq 
\der_0 A\, =\, \left\{ A, H \right\}^*, 
\label{16}
\feq 
defined in terms of the elementary brackets/anti-brackets
\beq 
\barr{ll} 
\left\{ C({\bf r},t), \pi_C({\bf r}^{\prime},t) \right\}^* = 
 \gd^3({\bf r} - {\bf r}^{\prime}), & \left\{ N({\bf r},t), 
 \pi_N({\bf r}^{\prime},t) \right\}^* = \gd^3({\bf r} - {\bf r}^{\prime}), \\
 \\
\left\{ \pi_C({\bf r},t), \bgps_{\pm}({\bf r}^{\prime},t) \right\}^* = \dsp{
 \frac{G^{\prime}(C)}{2G(C)}\, \bgps_{\pm}({\bf r},t)\, 
 \gd^3({\bf r} - {\bf r}^{\prime}), } & 
\left\{ \gps_{\pm}({\bf r},t), \pi_C({\bf r}^{\prime},t) \right\}^*\, = 
 \dsp{- \frac{G^{\prime}(C)}{2G(C)}\, \gps_{\pm}({\bf r},t)\, 
 \gd^3({\bf r} - {\bf r}^{\prime}), }\\ 
 \\
\left\{ \gps_{\pm}({\bf r},t), \bgps_{\mp}({\bf r}^{\prime},t) 
 \right\}^*\, = \dsp{ \frac{\gg^0 (1 \mp \gg_5)}{2G(C)}\, 
 \gd^3({\bf r} - {\bf r}^{\prime}). } &  
\earr
\label{17}
\feq  
In particular we note, that the equation 
\beq 
\dot{\pi}_N = \left\{ \pi_N, H_{eff} \right\}^* = \vec{\nabla} \cdot \left[ 
 \frac{1}{G(C)} \lh \vec{\nabla} N + \frac{i}{2}\, G^{\prime}(C)\, 
 \bgps_+ \vec{\gg}\, \gps_- \rh \right],
\label{18}
\feq 
after the identification (\ref{13}) is equivalent with $\dot{V}_0 = 
\vec{\nabla} \cdot \vec{V}$, or $\der \cdot V = 0$.
A somewhat tedious calculation shows that (\ref{16}) and (\ref{17}) indeed 
reproduce all covariant field equations (\ref{7}) and (\ref{8}). We can now 
construct the canonical expressions for the conserved four-momentum and the 
supercharges; for the energy-momentum vector we find the results
\beq 
P_0\, =\, H\, =\, \int d^3{\bf r}\, \cH, \hs{2} 
P_i\, =\, \int d^3{\bf r}\, \left[ \pi_C \nabla_i C + \pi_N \nabla_i N
 + \frac{1}{2}\, G(C)\, \bgps_+ \gg_0 \stackrel{\leftrightarrow}{\nabla}_i 
 \gps_- \right]. 
\label{20}
\feq 
The canonical expressions for the supercharges are 
\beq 
\barr{l} 
Q_+ = \dsp{ \int d^3{\bf r}\, \left[ \lh \pi_C - i\, G(C)\, 
 \pi_N \rh \gps_+ + \lh G(C)\, \nabla\Slashed C - 
 i \nabla\Slashed N \rh \gg_0 \gps_+ + \frac{1}{4}\, G^{\prime}(C)\,
 \gg_0 \gps_-\, \bgps_+ \gps_+ \right], }\\
 \\
Q_- = \dsp{ \int d^3{\bf r}\, \left[ \lh \pi_C + i G(C)\, 
 \pi_N \rh \gps_-  + \lh G(C)\, \nabla\Slashed C + 
 i \nabla\Slashed N \rh \gg_0 \gps_- + \frac{1}{4}\, G^{\prime}(C)\,
 \gg_0 \gps_+\, \bgps_- \gps_- \right]. }
\earr
\label{30}
\feq 
Like the hamiltonian generates the time-evolution, the supercharges generate 
the supersymmetry transformations: the results (\ref{3}) are reproduced by
the brackets 
\beq
\gd(\ge_{\pm})\, A = \left\{ A, \bge_{\pm} Q_{\pm} \right\}^*. 
\label{31}
\feq
The supercharges satisfy the standard super-Poincar\'{e} algebra, in 
particular $\left\{ Q_{\pm}, \bQ_{\mp} \right\}^* = 2 P\Slashed$. 
Note that the supersymmetry transformation rule implies $\go_{\mu} \equiv 
\left\{ V_{\mu}, Q_{\pm} \right\}^* =   2 i\, \gs_{\mu\nu} \der^{\nu}
\gps_{\pm}$, where $\go_{\mu}$ is a conserved fermionic current: $\der \cdot 
\go = 0$, as expected from supersymmetry and the conservation law of 
$V_{\mu}$. It is also of interest to discuss the brackets of the current 
components among themselves. By applying the identifications (\ref{13}) one 
finds the  non-trivial results 
\beq
\left\{ V_0({\bf r},t), \vec{V}({\bf r}^{\prime},t) \right\}^* = 
 \frac{1}{G(C)}\, \vec{\nabla}_{\bf r}\, \gd^3({\bf r} - 
 {\bf r}^{\prime}), \hs{1} \mbox{and} \hs{1}
\left\{ V_i({\bf r},t), V_j({\bf r}^{\prime},t) \right\}^* = 
 \frac{\lh G^{\prime}(C) \rh^2}{\lh G(C) \rh^3}\,\, 
 \bgps_+ \gg^0 \gs_{ij} \gps_-\,\, \gd^3({\bf r} - {\bf r}^{\prime}). 
\label{34}
\feq 
Note that the charge-conjugation properties of the spinors 
guarantee that $\bgps_+ \gg^0 \gs_{ij} \gps_- = i \gps^{\dagger}_- \gs_{ij} 
\gps_- = i \gps^{\dagger}_+ \gs_{ij} \gps_+$, as in our conventions 
$\gs_{ij}$ is anti-hermitean. \nl

For easy comparison we have performed the canonical analysis in terms of the 
fields related by the linear supersymmetry transformations (\ref{3}), at the 
price of dealing with off-diagonal terms in the Dirac-Poisson brackets 
(\ref{17}). Observe, however, that the brackets can be diagonalized by a 
field redefinition 
\beq 
\gF_{\pm} = \sqrt{G(C)}\, \gps_{\pm}. 
\label{35}
\feq 
Indeed, in terms of the new fermion fields the brackets read 
\beq 
\barr{l} 
\left\{ C({\bf r},t), \pi_C({\bf r}^{\prime},t) \right\}^* = 
 \gd^3({\bf r} - {\bf r}^{\prime}), \hs{2} \left\{ N({\bf r},t), 
 \pi_N({\bf r}^{\prime},t) \right\}^* = \gd^3({\bf r} - {\bf r}^{\prime}), \\
 \\
\left\{ \pi_C({\bf r},t), \bgF_{\pm}({\bf r}^{\prime},t) \right\}^* = 
\left\{ \gF_{\pm}({\bf r},t), \pi_C({\bf r}^{\prime},t) \right\}^*\, = 0, \\ 
 \\
\left\{ \gF_{\pm}({\bf r},t), \bgF_{\mp}({\bf r}^{\prime},t) 
 \right\}^*\, = \dsp{ \frac{1}{2}\, \gg^0 (1 \mp \gg_5)\, 
 \gd^3({\bf r} - {\bf r}^{\prime}), }  
\earr
\label{36}
\feq  
but of course the supersymmetry transformations (\ref{3}) now become 
non-linear. We do not present here the new expressions for the energy-momentum 
and supercurrents; however, the conserved current (\ref{1}) is of particular 
interest and we observe that after the field redefinition (\ref{35}) its 
space components read 
\beq 
\vec{V} = \frac{1}{G(C)}\, \vec{\nabla} N + \frac{i 
 G^{\prime}(C)}{2 (G(C))^2}\, \bgF_+ \vec{\gg}\, \gF_-. 
\label{37}
\feq 
To complete its hydrodynamical interpretation, we relate the fields in our 
model to the fluid density $\gr$ and velocity $u_{\mu}$. First we consider 
the bosonic reduction, obtained by requiring the fermion field to vanish: 
$\gF_{\pm} = 0$. The field equations (\ref{7}) then reduce to 
\beq 
\barr{l} 
\dsp{ V_{\mu} = \gr u_{\mu} = \frac{1}{G(C)}\, \der_{\mu} N,
\hs{2}  V_{\mu}^2 = -\gr^2, \hs{2} \der \cdot V = 0, }\\  
 \\ 
\dsp{ G(C) \Box\, C = \frac{1}{2} G^{\prime}(C) \lh - (\der C)^2 + \gr^2 \rh.} 
\earr 
\label{40} \label{41} \label{42} 
\feq 
In the present context the current-conservation condition is to be interpreted
as the equation of continuity in hydrodynamics, whilst the last equation
(\ref{40}) can be used to express the density $\gr$ in terms of the field $C$. 
The first equation (\ref{40}) then implies that the velocity field for fixed 
$\gr$ (i.e.\ for fixed $C$) depends only on one degree of freedom, $N$. This 
is also the situtation encountered in potential flow, when $u_{\mu} = 
\der_{\mu} \gth$, $(\der \gth)^2 = -1$. Therefore our bosonic model describes
potential flow with vanishing vorticity, provided
\beq 
\der_{\mu} \gth = \frac{1}{\gr G(C)}\, \der_{\mu} N. 
\label{43}
\feq 
Such a relation is consistently realized if either $\der_{\mu} \gr \sim 
\der_{\mu} \gth \sim V_{\mu}$, or else if $N$ can be expressed entirely in
terms of $\gth$, independend of $\gr$: $\der N/\der \gr = 0$. In the first 
case $\der_{\mu} C \sim V_{\mu}$, and the bosonic part of the energy-momentum
tensor manifestly takes the  perfect fluid form; indeed, if $\der_{\mu} C = 
\gk u_{\mu}$, then
\beq
T_{\mu\nu}^{(B)} = p g_{\mu\nu} + (p + \gve) u_{\mu} u_{\mu}, 
\label{43.1}
\feq 
where the pressure and energy density are given by\footnote{It is possible
to modify the action (\ref{6}) so as to add a constant term to the energy; 
then the equation of state is changed to $p + \mu^2 = \gve - \mu^2 =
\frac{1}{2}\, G(C) \lh \gk^2 + \gr^2 \rh$; details are given in \ct{jwtnsgn}.} 
$p = \gve = \frac{1}{2}\, G(C) \lh \gk^2 + \gr^2 \rh$. The case $\der
N/\der \gr = 0$ is realized if $N = 2m \gth$, $\gr = 2m/G(C)$, with $m$ a
constant of proportionality. It follows that $N$ must satisfy the condition
$(\der N)^2 = - 4m^2$ = constant, whilst combination of these conditions with 
(\ref{42}) gives 
\beq 
\frac{[G(C)]^{5/2}}{G^{\prime}(C)}\, \der^{\mu} \lh \sqrt{G(C)}\, \der_{\mu}
C \rh = 2 m^2.  
\label{45} 
\feq 
For example, if $G(C) = 1/C$, we find the imaginary-mass Klein-Gordon equation
\beq 
\lh \Box + m^2 \rh \sqrt{C} = 0,
\label{46}
\feq 
which allows non-trivial stationary solutions for the density $\gr = 2mC$. 
The more general Ansatz $G(C) = C^p$ leads to the non-linear relation 
\beq 
\Box\, C^{(p+2)/2}\, =\, m^2\, p (p+2)\, C^{-(3p+2)/2}. 
\label{47}
\feq 
If the fermion fields are switched on, the potential character of the 
flow disappears; indeed,  as 
\beq 
V_{\mu} = \gr u_{\mu} = \frac{1}{G(C)}\, \der_{\mu} N + 
 \frac{i G^{\prime}(C)}{2(G(C))^2}\, \bar{\gF}_+ \gg_{\mu} \gF_-, 
\label{48}
\feq 
the velocity is now parametrized in terms of one bosonic and two fermionic 
degrees of freedom. As eq.(\ref{34}) shows, the vorticity is non-zero
as a  result of the fermionic contribution indeed. In terms of the redefined 
fields $\gF$ these bracket relations take the form 
\beq
\left\{ V_i({\bf r},t), V_j({\bf r}^{\prime},t) \right\}^* = 
 \lh \frac{G^{\prime}(C)}{[G(C)]^2}\rh^2\, \bgF_+ \gg^0 \gs_{ij} \gF_-\,\, 
 \gd^3({\bf r} - {\bf r}^{\prime}). 
\label{49}
\feq 
In particular, for the case $G(C) = 1/C$ the coefficients of the bi-fermion 
term on the r.h.s.\ of eqs.(\ref{48}) and (\ref{49}) reduce to constants, 
and the current is a linear combination of free boson and fermion currents. 

The models constructed here can be generalized to include vorticity from 
bosonic potentials, by a supersymmetric extension of the Clebsch decomposition 
of $V_{\mu}$; this is explained elsewhere \ct{jwtnsgn}. Consistent
extensions of the models at the quantum level are of potential interest in
cosmology, where they could provide an effective description of an early
supersymmetric phase of the universe, and in condensed matter physics, where
they might apply to quantum fluids like $^3$He-$^4$He mixtures, up to
effects proportional to the mass-differences of these isotopes.

\end{document}